\documentclass[fleqn,usenatbib]{mnras}

\usepackage{newtxtext,newtxmath}
\usepackage[T1]{fontenc}

\DeclareRobustCommand{\VAN}[3]{#2}
\let\VANthebibliography\thebibliography
\def\thebibliography{\DeclareRobustCommand{\VAN}[3]{##3}\VANthebibliography}

\usepackage{graphicx}	% Including figure files
\usepackage{amsmath}	% Advanced maths commands
\usepackage{amssymb}	% Extra maths symbols

\newcommand{\ergpcms}{\,erg\,cm$^{-2}$\,s$^{-1}$}
\newcommand{\msun}{\,M$_{\mathrm{\odot}}$}
\newcommand{\sax}{SAX~J1808.4$-$3658}

\title[2019 outburst rise of SAX J1808.4-3658]{Enhanced optical activity 12 days before X-ray activity, and a 4 day X-ray delay during outburst rise, in a low-mass X-ray binary}
\author[A. J. Goodwin et al.]{A. J. Goodwin,$^{1}$\thanks{E-mail: ajgoodwin.astro@gmail.com}
D. M. Russell,$^{2}$
D. K. Galloway,$^{1}$
M. C. Baglio,$^{2,3}$
A. S. Parikh,$^{4}$
\newauthor
D. A. H. Buckley,$^{5}$
J. Homan,$^{6,7}$
D. M. Bramich,$^{2}$
J. J. M. in 't Zand,$^{6}$
C. O. Heinke,$^{8}$
\newauthor
E. J. Kotze, $^{5,9}$
D. de Martino, $^{10}$
A. Papitto,$^{11}$
F. Lewis,$^{12,13}$
and 
R. Wijnands$^{4}$
\\
$^{1}$School of Physics and Astronomy, Monash University, Clayton, 3800, Australia\\
$^{2}$Center for Astro, Particle and Planetary Physics, New York University Abu Dhabi, PO Box 129188, Abu Dhabi, UAE\\
$^{3}$INAF, Osservatorio Astronomico di Brera, Via E. Bianchi 46, I-23807 Merate (LC), Italy \\
$^{4}$Anton Pannekoek Institute for Astronomy, University of Amsterdam, Postbus 94249, NL-1090 GE Amsterdam, The Netherlands\\
$^{5}$South African Astronomical Observatory, PO Box 9, Observatory 7935, Cape Town, South Africa\\
$^{6}$Eureka Scientific, Inc., 2452 Delmer Street, Oakland, CA 94602, USA\\
$^{7}$SRON Netherlands Institute for Space Research, Sorbonnelaan 2, 3584 CA, Utrecht, The Netherlands \\
$^{8}$Department of Physics, University of Alberta, CCIS 4-183, Edmonton, AB T6G 2E1, Canada\\
$^{9}$Southern African Large Telescope Foundation, PO Box 9, Observatory 7935, Cape Town, South Africa\\
$^{10}$INAF – Osservatorio Astronomico di Capodimonte, Salita Moiariello 16, 80131 Napoli, Italy\\
$^{11}$INAF – Osservatorio Astronomico di Roma, via Frascati 33, 00040 Monte Porzio Catone (Roma), Italy\\
$^{12}$Faulkes Telescope Project, School of Physics and Astronomy, Cardiff University, The Parade, Cardiff, CF24 3AA, Wales, UK \\
$^{13}$Astrophysics Research Institute, Liverpool John Moores University, 146 Brownlow Hill, Liverpool L3 5RF, UK
}

\date{Accepted 2020 August 20. Received 2020 August 20; in original form 2020 June 4}

\pubyear{2020}

\begin{document}
\label{firstpage}
\pagerange{\pageref{firstpage}--\pageref{lastpage}}
\maketitle

% Abstract of the paper
\begin{abstract}

X-ray transients, such as accreting neutron stars, periodically undergo outbursts, thought to be caused by a thermal-viscous instability in the accretion disk. Usually outbursts of accreting neutron stars are identified when the accretion disk has undergone an instability, and the persistent X-ray flux has risen to a threshold detectable by all sky monitors on X-ray space observatories. Here we present the earliest known combined optical, UV, and X-ray monitoring observations of the outburst onset of an accreting neutron star low mass X-ray binary system. 
We observed a significant, continuing increase in the optical $i'$-band magnitude starting on July 25, 12 days before the first X-ray detection with \textit{Swift}/XRT and NICER (August 6), during the onset of the 2019 outburst of SAX J1808.4--3658. We also observed a 4 day optical to X-ray rise delay, and a 2 day UV to X-ray delay, at the onset of the outburst. We present the multiwavelength observations that were obtained, discussing the theory of outbursts in X-ray transients, including the disk instability model, and the implications of the delay. This work is an important confirmation of the delay in optical to X-ray emission during the onset of outbursts in low mass X-ray binaries, which has only previously been measured with less sensitive all sky monitors. We find observational evidence that the outburst is triggered by ionisation of hydrogen in the disk.

\end{abstract}

% Select between one and six entries from the list of approved keywords.
% Don't make up new ones.
\begin{keywords}
accretion, accretion discs -- X-rays: binaries -- X-rays: individual: SAX J1808.4-3658  
\end{keywords}

%%%%%%%%%%%%%%%%%%%%%%%%%%%%%%%%%%%%%%%%%%%%%%%%%%

%%%%%%%%%%%%%%%%% BODY OF PAPER %%%%%%%%%%%%%%%%%%
%Specific main argument: ``We observed a 12 day delay between the optical and X-ray emission of an accreting neutron star coming into outburst, which coincides with the theoretical viscous timescale of the accretion disk"

\section{Introduction}\label{sec:introduction}
% * opening
%     * importance of research area
%     * background information
%     * current research focus
% * development
%     * overview of recent works
%     * gap in research
%     * specific problem to be addressed
% * closing
%     * overview of current work
%much of this section should focus on the findings of studies that looked at similar objectives. 

% Introduce general topic, context:
%accreting neutron stars and outbursts:
Transient accreting neutron stars in low mass X-ray binaries (LMXBs) are in close binary orbits with a main sequence, white dwarf, or other low mass star. The neutron star accretes from the companion via Roche Lobe overflow, forming an accretion disk, and emitting energetic radiation in the form of X-rays when material transfers from the accretion disk onto the neutron star \citep[e.g.][]{White1985,Lewin2006}. Outbursts in transient accreting neutron stars are characterised by an abrupt increase in X-ray luminosity of several orders of magnitude over a few days, typically followed by a decay on a timescale of a month or several months, before the system returns to its low luminosity quiescent state \citep[e.g.][]{Frank1987}.  

% different kinds of emission:
Optical emission in these systems is generally thought to arise from the outer accretion disk, companion star, and at higher luminosity X-ray reprocessing in the disk, in some cases with a contribution from synchrotron emission from jets \citep*[e.g.][]{Russell2007}. Whereas X-rays are thought to originate from close to the neutron star, at the inner disk from the hot accretion flow \citep{Lasota2001}, or from the neutron star surface. Thus, we can use multiwavelength observations of LMXBs to follow the progression of an outburst, and separate the activity of the accretion disk, jet, companion star, and neutron star. 

% 1808:
SAX J1808.4--3658 was the first accretion-powered millisecond pulsar (AMSP) to be discovered \citep{Zand1998,Wijnands1998}, and %consistently 
goes into outburst %approximately 
every few years \citep{Bult2019,DelSanto2015,Zand2013,Markwardt2011,Markwardt2008,Markwardt2002,Wijnands2001}. The outbursts of SAX J1808.4--3658 have  been occurring with progressively longer recurrence times \citep[e.g.][]{Galloway2006}, which is a puzzling phenomenon. The system consists of a 401 Hz pulsar \citep{Wijnands1998} in a close binary orbit \citep[P=2.01\,hr;][]{Chakrabarty1998}, with a very low mass \citep[$M=0.03-0.11$\msun;][]{Deloye2008,Elebert2009,Wang2013}
companion star that is likely hydrogen depleted \citep{GoodwinMCMC,Johnston2018,Galloway2006}. During outburst, the system exhibits type-I thermonuclear X-ray bursts \citep[e.g.][]{Galloway2006} with burst oscillations at the pulsar frequency \citep{Zand2001,Chakrabarty2003}.

% Map previous research, moving from general to more specific:
% The DIM:
The exact mechanism that causes the onset of an outburst in LMXBs is unknown. The disk instability model (DIM) is the prevalent theory that describes a mechanism for outburst: a thermal-viscous disk instability \citep[see][for a review]{Dubus2001}. The theory was first proposed in the 1970s to explain similar outbursts observed in dwarf novae systems \citep[a subset of cataclysmic variables (CVs);][]{Smak1971, Hoshi1979, Osaki1974}. The DIM was extended to encompass LMXBs around 20 years later when the similarity in the fast-rise and exponential-decay behaviour between outbursts in CVs and transient LMXBs was recognised \citep{vanParadijs1984,Cannizzo1985,vanParadijs1996}. There have been some significant and important modifications to the DIM necessary to explain the variations in, and general observational behaviour of, LMXBs. These include the effects of irradiation of the accretion disk and the donor star by the LMXB primary; inner disk truncation, if the compact object has a strong ($>10^8$\,G) magnetic field or due to evaporation; variations in mass transfer; and the existence of winds and outflows and the torque they exert on the accretion disk \citep[e.g.,][]{Hameury2019}. Nevertheless, the single underlying mechanism that describes the cause of outburst in an accretion disk has remained constant since the theory was first proposed; a thermal-viscous disk instability. 

According to the DIM, the outburst-quiescence cycle of an LMXB may proceed as follows: first, during quiescence, the cold disk accumulates mass via Roche-lobe overflow from the companion until the disk reaches a critical density, and the disk temperature rises to the critical value required to ionise hydrogen. This event causes a heating front to propagate through the disk, bringing it to a hot, bright state, and commencing the outburst \cite[e.g.][]{Menou1999}. The outburst lasts on the order of a month, during which accretion onto the compact object gives rise to high energy X-ray emission. The outburst then decays in luminosity as the disk depletes, and the system returns to quiescence to begin building up mass in the disk for the next outburst \citep[e.g.][]{Lasota2001}. This picture is a very simplified rendition of what is happening during an outburst of an LXMB, and there is variation observed in the behaviour of many systems. One such key variation includes the time delay between when the disk instability first occurs, and the heating front begins to propagate in the disk; and when the outburst commences, and accretion onto the compact object is observed as an increase in X-ray luminosity. 

\citet{Dubus2001} describe two different kinds of outburst, depending on how long it takes for the heating front to propagate in to the inner disk. Two heating fronts are formed at the point of ignition in the disk, and they propagate both inwards and outwards from the ignition radius \citep{Menou1999}. For ``inside-out" outbursts, the ignition occurs at a small radius, and the inward propagating front reaches the inner disk much faster than the outward propagating front reaches the outer disk. In ``outside-in" outbursts, the ignition occurs at a larger disk radius, and it can take much longer for the heating front to reach the inner disk. 
\citet{Dubus2001} describe the primary difference between ``inside-out" and ``outside-in" outbursts.  During quiescence, the density profile of the disk is approximately linearly proportional to its radius, which implies that ``outside-in" heating fronts always progress through regions of decreasing surface density. Whereas, in ``inside-out" outbursts, the outward front can encounter regions of higher densities, and if the heating front cannot raise the density of the front above the higher density, it stalls and a cooling front can develop. ``Inside-out" outburst fronts propagate slowly, and it can take much longer to bring the disk to the hot state than in ``outside-in" outbursts. When irradiation is accounted for, it does not change the structure of the heating front, but it does change the maximum radius to which an ``inside-out" outburst can propagate. The ignition radius depends on the mass transfer rate from the secondary and the size of the disk. \citep{Smak1984,Hameury1998}. For low mass transfer rates, the accreted matter will drift inwards in the cold disk, and the maximum surface density will accumulate in the inner disk, giving rise to inside-out outbursts. If the mass transfer rate is high, the mass accumulation time at the outer radius can become lower than the drift time, and an outside-in outburst will be triggered by the higher surface density in the outer disk. \citet{Dubus2001} found that the mass transfer rate required for an outside-in outburst to occur in their model of a soft X-ray transient was higher than the accretion rate for which the disk was stable, thus indicating that inside-out outbursts are more likely for this kind of system. However, an outside-in outburst is theoretically possible if the accretion disk has a small radius.

%Past observations of delays/rises:
Observations, especially over multiple wavelengths, of the rise to outburst in LMXB systems are rare, as the rise occurs rather quickly (on the order of a few days), and outbursts are not usually detected until the flux has risen above the detection threshold of all-sky monitors aboard X-ray space observatories. The X-ray Binary New Early Warning System \citep[XB-NEWS;][]{Russell2019} was developed with the goal of producing light curves in real time for long-term optical monitoring of LMXBs \citep[][and other sources]{Lewis2008} with Las Cumbres Observatory network telescopes. One of the main aims of this pipeline is to be able to identify new outbursts of LMXBs from an early optical rise, in real time, which will increase the number of sources that have observationally constrained optical to X-ray delays. There are a few systems in which the optical to X-ray delay has been constrained using observations from X-ray all sky monitors. These include: a $<7$\,d delay in V404 Cyg \citep{Bernardini2016}, a $<6$\,d delay in GRO J1655--40 \citep{Orosz1997,Hameury1997}, a $<9$\,d delay in XTE J1550--564 \citep{Jain2001}, a $<10$\,d delay in XTE J1118+480 \citep{Wren2001,Zurita2006}, a $<5$\,d delay in 4U 1543--47, and a $<7$\,d delay in ASASSN--18ey \citep{Tucker2018}, which are all black hole low mass X-ray binaries. There is one neutron star X-ray binary, Aql X--1, in which a $<3$--8\,d optical to X-ray delay has been inferred \citep{Shahbaz1998,Russell2019}, but again this constraint was only obtained with an X-ray all sky monitor. Prior to this work, to date, there are no observations of the rise to outburst of an LMXB that conclusively constrain the delay time using an X-ray telescope more sensitive than an all sky monitor.

% This study description:
In this work we present the earliest multiwavelength observations of the rise to outburst of the accreting neutron star SAX J1808.4--3658. In Section \ref{sec:method}, we outline the observations and data processing of the 2019 outburst we obtained. In Section \ref{sec:results} we present the multiwavelength light curve, including an optical spectrum obtained on the day of the first X-ray detection, and the disk temperature evolution inferred from the optical $V$-$i'$ color. In Section \ref{sec:discussion}, we discuss the implications of a 12 day delay between the first optical and X-ray detections of the source, and relate these observations to the theoretical expectations from the disk instability model.

\section{Methods and Observations}\label{sec:method}
%You need to provide enough information for others to understand exactly what you did (and where the decision is important, why) and to repeat the experiments.

We monitored the accretion-powered millisecond pulsar SAX J1808.4--3658 during the lead-up to its 2019 outburst using the Neil Gehrels Swift Observatory (\textit{Swift}), the 2-m Faulkes Telescope South (at Siding Spring, Australia), the Las Cumbres Observatory (LCO) network of 1-m robotic telescopes, and the South African Large Telescope (SALT). 

% suggested addition from dkg
Our monitoring program was motivated by the prediction of a phenomenological (quadratic) model for the times of the previous 8 outbursts, since the first in 1996. The model predicted the next outburst in 2019 May, but no activity was seen either in that month or the following.
We began our observing program on 2019 July 17, and observed a statistically significant ($>5$-$\sigma$) increase from the quiescent level in the optical $i'$-band magnitude on July 25 (MJD 58689). We 
% did not detect the source in 
first detected X-ray activity with the \textit{Swift} X-ray Telescope (XRT) 
% band until 
on August 6.44 \cite[MJD 58701.44;][]{goodwin19}, 12 days after the first sign of increased optical activity. The X-ray flux subsequently increased over the next 2 days, indicating the beginning of accretion onto the neutron star, and the onset of outburst. 

% additional para from dkg
The 2019 outburst occurred 85~d later than the time predicted by the phenomenological model, after a quiescent interval of 4.3~yr. This interval is the longest separating any two outbursts of \sax, by almost a year. The outburst interval has increased after every outburst, up from 1.6~yr between the first two, in 1996 September and 1998 April. As noted by \cite{gal08b}, the continuing trend of increasing outburst interval contributed to a decline in the long-term averaged accretion rate. Provided the quadratic model holds for the next outburst, we expect it to occur no earlier than MJD 60260 (2023 November 12), another 4.3~yr after the 2019 outburst. The likely uncertainty on this prediction is as for the present outburst, $\approx90$~d.

We  present below a description of the optical observations, as well as the X-ray observations with \textit{Swift} and 
% the X-ray lightcurve from 
the Neutron Star Interior Composition Explorer (NICER) instrument; the NICER observations have already been reported in \citet{Bult2019}. 

\subsection{Optical observations}
\subsubsection{LCO monitoring}

SAX J1808.4–3658 was extensively monitored throughout the 2019 outburst from the very beginning with the Las Cumbres Observatory (LCO) network including the 2-metre Faulkes Telescope South. 
The LCO observations are part of an ongoing monitoring campaign of$\sim$50 low-mass X-ray binaries \citep{Lewis2008} co-ordinated by the Faulkes Telescope Project. Observations were performed with the $B$, $V$, $R$ Bessell filters and the SDSS $u', i', z'$ filters (354--870\,nm). For the purposes of this paper, we include the $B$, $V$, $R$, $i'$, and $z'$-band measurements of the first part of the outburst. The full data set in all filters will be  published in Baglio et al. (submitted) and Russell et al. (in preparation). Aperture photometry was performed by a new data analysis pipeline, the ``X-ray Binary New Early Warning System (XB-NEWS)'', developed with the aim of automatically analysing any image of a specific list of targets that is acquired with the LCO network telescopes, including new images in real-time. The pipeline first computes an astrometric solution on each image using Gaia DR2\footnote{\url{https://www.cosmos.esa.int/web/gaia/dr2}} positions, then performs accurate aperture photometry of all the stars in the field of view. Zero-point calibrations between epochs are solved for using the method described in \citet{Bramich2012}, and finally the pipeline flux calibrates all the stars on a standard magnitude scale using the ATLAS-REFCAT2 catalogue \citep{Tonry2018}\footnote{\url{https://archive.stsci.edu/prepds/atlas-refcat2/}} (which includes PanSTARRS DR1, APASS, and other catalogues). XB-NEWS also performs multi-aperture photometry (similar to the DAOGROW algorithm by \citealt{Stetson1990}) which is effectively azimuthally-averaged PSF profile fitting photometry. This method is optimised for point sources in fields that are not too crowded, and it was used to extract the magnitudes of SAX J1808.4--3658 for this paper. 

For images in which the source was not formally detected above the detection threshold, forced multi-aperture photometry at the known location of the source was performed, and magnitudes with errors $> 0.25$ mag were excluded as these are only very marginal detections, or non-detections. SAX J1808.4--3658 lies in a very crowded field of the Galactic plane, with a few stars within 1$^{\prime\prime}$ of the source position \citep[e.g.][]{Deloye2008}. These stars can contribute to the quiescent flux measurements, but this does not affect the date of the early rise into outburst since these stars are unchanging and the rise does not correlate with seeing. 

The result of the pipeline process is a calibrated light curve for each object in the field of view. For further details see \citet{Russell2019,Pirbhoy2020}. We also extinction corrected the $V$ and $i'$ magnitudes using $A_V=0.510$\,mag, where we have inferred $A_V$ given the relationship between $N_{\mathrm{H}}$ and $A_V$ for our Galaxy, reported in \citet{Foight2016} and using the measured $N_{\mathrm{H}}$ for the 2019 outburst of SAX J1808.4--3658 of $(1.5 \pm 0.1)\times10^{21}$\,cm$^{-2}$ \citep{DiSalvo2019}. Using the \citet{Cardelli1989} extinction law, we infer $A_{i'}= 0.340$\,mag.

The LCO monitoring of SAX J1808.4--3658 was crucial to detect the first flux enhancement; the early precursor of the new outburst, on July 25 \citep[MJD 58689.48;][]{Russell2019ATel12964} in $i'$-band.

\subsubsection{SALT Spectroscopy}
We took optical spectroscopic observations of SAX J1808.4--3658 with the Southern African Large Telescope \citep[SALT;][]{Buckley2006SPIE.6267E..0ZB}, on August 2, 3 and 6 (MJD 58697, 58698, and 58701). We covered the region 4060--7120\,\AA{} at a mean resolution of 4.7\,\AA{} using the Robert Stobie Spectrograph \citep[RSS;][]{Burgh2003SPIE.4841.1463B} with the PG900 VPH grating in frame transfer mode with 200\,s exposures and a slitwidth of 1.5$^{\prime\prime}$. No spectral trace was detected on August 2 in an accumulated 2400\,s exposure, although there was a focus issue. A weak spectral trace was detected on August 3 in an accumulated 2200\,s exposure. On August 6 a spectrum of the object was obtained with 12 repetitions of 200\,s exposures, revealing Balmer absorption lines and strong interstellar NaD absorption. The seeing on August 6 was 1.8$^{\prime\prime}$. We also obtained follow up spectra on August 22, 25, and September 26 (MJD 58717, 58720, and 58752), which will be presented in a follow up paper (Russell et al., in preparation).

%\A{David B to fill in how the SALT spectrum on Aug 6 was processed}
%- image reduction: bias subraction? Flat correction? 
%- spectrum extraction: software package?
%- wavelength and flux calibration

The spectroscopic
data reductions were done using the PySALT version 0.47, the PyRAF-based software package for SALT data reductions
\citep{Crawford2010SPIE.7737E..25C}\footnote{\url{https://astronomers.salt.ac.za/software/pysalt-documentation/}}, which includes gain and amplifier cross-talk corrections, bias subtraction, amplifier
mosaicing, and cosmetic corrections. Spectral reductions (object extraction, wavelength calibration and background subtraction) were all done using standard IRAF\footnote{\url{https://iraf.noao.edu/}} routines, as was the relative flux calibration, 
%which is only 
as absolute flux calibration is not 
possible with SALT due to its unusual design \citep{10.1093/mnrasl/slx196}. Low frequency bumps are artefacts from the imperfect flux calibration.

\subsection{\textit{Swift} observations}

\subsubsection{XRT}
The XRT aboard \textit{Swift} is sensitive in the 0.2--10\,keV energy range, has an effective area of 100\,cm$^{2}$, a 23.6$\times$23.6$^{\prime}$ field of view, and an 18$^{\prime\prime}$ resolution  at half-power diameter \citep{Burrows2005}. We obtained a total of 14 observations between July 21 and August 9 (MJD 58685 to 58705), with an average exposure time of 0.5\,ks, and a cadence of 2 days. The 8 observations between July 21 and August 4 were carried out with the XRT in Windowed Timing (WT) mode, the observation on August 6 with the XRT in Photon Counting (PC) mode, and the remaining observations from August 8 with the XRT in Auto mode. WT mode provides 1 dimensional imaging at the orientation of the roll angle of the spacecraft, with a sensitivity limit of approximately 2.4$\times10^{-11}$\,erg\,cm$^{-2}$\,s${^{-1}}$. Photon counting mode permits full spectral and spatial information to be obtained for sources, capable of accurately measuring fluxes down to $2\times10^{-14}$\,erg\,cm$^{-2}$\,s${^{-1}}$ in $10^4$\,s. To extract the X-ray flux measured by the XRT, we examined all 14 observations (OBSID 00030034120--00033801025, July 21--August 9) using the \textit{Swift} online XRT product builder \citep{Evans2009}. The observations prior to Aug 6 (OBSID 00030034120--00030034127) could not be processed with the online XRT product builder, so we manually extracted upper limits using a box of 20 by 5\,pixels (aligned to match the data) to extract the source flux, and a larger region (typically $>40$\,pixels by $5$\,pixels) to extract the background. In none of these observations was the source clearly visible to the eye. Exposure maps and ancillary response functions were generated, with the appropriate RMF file from the \textit{Swift} calibration website. The upper limits obtained are 90$\%$ (two-sided) confidence levels. Two instances where the FOV of \textit{Swift} missed the target (OBSID 00030034122 and 00030034128), and one instance where the observation was only 40\,s long (OBSID 00030034128) were excluded.

\subsubsection{UVOT}
The UltraViolet and Optical Telescope (UVOT) aboard \textit{Swift} is a 30\,cm Ritchey-Chrétien reflector with a 17$\times$17$^{\prime}$ field of view, sensitive to wavelengths in the range 1700--6500\,\AA{}, and operates in photon counting mode. It has 7 filters, 4 of which were used in this project. The UVW1 filter has a peak sensitivity at 2600\,\AA{}, the UVW2 filter at 1928\,\AA{}, the UVM2 filter at 2245\,\AA{} and the U filter at 3465\,\AA{} \citep{Roming2005}. To extract the UV flux for each filter we examined the observations between July 19 and August 10 (MJD 58683 to 58705). We re-aligned the images, used a circular source region of 5$^{\prime\prime}$ and background region consisting of 3$\times$5$^{\prime\prime}$ apertures (due to the field being very crowded in UV it is not possible to find a bigger background region). We used the HEAsoft Swift software tools\footnote{\url{https://heasarc.nasa.gov/lheasoft/}} \textsc{uvotsource} task to carry out the aperture photometry and construct the light curve. The error bars and upper limits correspond to 1-$\sigma$. Two instances of very low exposure images leading to unconstrained upper limits have been rejected (OBSID 30034125 with the UVW2 filter, the only filter used during this observation, and 00030034140 with the UVW1 filter, one of three filters used during this observation with the others being UVM2 and UVW2). To extinction correct the UV magnitudes we used the relation with $A_V$ described in \citet{Mathis1990} to infer that $A_U=0.82$, $A_{W1}=1.08$, $A_{M2}=1.49$, $A_{W2}=1.36$. %{\tt\bf (wouldn't uncertainties be appropriate here? - jz)}

\subsection{NICER observations}
The NICER X-ray Timing Instrument \citep{Gendreau2016} consists of 56 concentrators that are each coupled to a silicon drift detector that is housed in a Focal Plane Module (FPM). At the time of the SAX J1808.4--3658 observations, 52 of the 56 FPMs were functional, providing an effective area of $\sim$1750 cm$^2$ in the 0.2--12 keV band. We analysed 69 {\it NICER} ObsIDs of SAX J1808.4--3658 (2050260101--2584014301), corresponding to observations made between July 30 and October 13. The source was often observed multiple times per day, resulting in dense coverage of the pre-outburst phase and the rise.

We used the {\tt nicerl2} task in HEAsoft version 6.27.2 to reprocess all observations, applying gain solution `optmv7he' and using the default filter criteria. 
% Spectra were extracted 
We extracted spectra from all active FPMs, except $\#$14 and $\#$34, which often show excessive noise; 
% This was done 
and for all good-time-intervals (GTIs) 
% that had a length 
longer than 100~s. Since NICER is a non-imaging instrument, the background needs to be estimated. 
% For this 
We used the tool {\tt nibackgen3C50} (R.\,Remillard et al, in prep.)
% . Model 
which predicted background rates 
% were typically 
generally below 3 counts\,s$^{-1}$ in the 0.5--10 keV band. The extracted spectra were grouped to a minimum of 25 counts per bin. They were background subtracted and analysed in the 0.5-10 keV band with XSPEC 12.11.0 \citep{ar1996}, with the appropriate response files for gain solution `optmv7he', and using $\chi^2$ statistics. The extracted-spectra and model background spectra were also used to construct a background-subtracted 0.5--10 keV light curve, with one data point per GTI (pink symbols in Figure \ref{fig:lightcurve}). 

The first clear detection of SAX J1808.4--3658 with NICER during the rise was made on August 6, 22:00 UTC, when the net (background-subtracted) count rate was 9.16$\pm0.16$ counts\,s$^{-1}$. We fit the spectrum of that GTI with an absorbed power-law ({\tt tbnew $\times$ pegpwrlw} in XSPEC), with the photoionization cross section set to {\tt vern} and the abundances set to {\tt wilm}. Our best fit ($\chi^2$=88 for 108 degrees of freedom) yields an $N_{\rm H}$ of 0.33$\pm$0.03 atoms\,cm$^{-2}$, a power-law photon index of 2.43$\pm$0.07, and an unabsorbed 0.5--10 keV flux of (2.91$\pm$0.11)$\times10^{-11}$ erg\,cm$^{-2}$\,s$^{-1}$.

A potential detection was made one GTI earlier, on August 5 22:46 UTC, when the net count rate (1.44$\pm$0.06 counts\,s$^{-1}$) was higher than during all observations preceding it (maximum of 0.85 counts\,s$^{-1}$). However, %a further analysis showed
comparison to contemporaneous background measurements indicated
 that the flux was still background dominated, and there was no statistically significant detection.

Between July 30 and August 5 the net count rates varied between -0.22 and 1.44 counts\,s$^{-1}$, with an average of $0.402\pm0.003$ counts\,s$^{-1}$. This 
% average count 
rate is significantly above zero, possibly due to contaminating sources in NICER's $\sim$30 arcmin$^2$ field-of-view. The variations in the pre-outburst net count rate are likely due to limitations of the background model (which is still being refined). The 1-$\sigma$ standard deviation in the net count rate before the first detection is 0.32 counts\,s$^{-1}$. We consider the average pre-outburst net count rate plus three times the 1-$\sigma$ standard deviation, 1.37 counts\,s$^{-1}$, to be a good general estimate of the NICER detection limit of SAX J1808.4--3658. Using the best-fit model to the spectrum of the first detection, this count rate corresponds to an unabsorbed 0.5--10 keV flux of $\sim4\times10^{-12}$ erg\,cm$^{-2}$\,s$^{-1}$.

%A first possible detection was made on August 05  22:46 UTC (1.44$\pm$0.06 counts\,s$^{-1}$), but a spectral analysis suggests that the spectrum is still background dominated. We consider the detection in the following GTI, on August 6 22:00 UTC , to be the first secure detection of SAX J1808.4--3658 with {\it NICER} during the rise. 

\section{Results}\label{sec:results}
%results are the defining feature of any paper. Present the core of your work and describe the most important features. Should be presented as a clear and succinct narrative. 
% General format:
% [link to previous paragraph? To [aim], we [method]. 
% We found [key feature of results (fig/table ref). [Key feature of results (fig/table ref)]. [Key feature of results (fig/table ref)]. This suggests [conclusion from these results]. 
% start the next paragraph with an aim that builds on the take-home-message, thus gradually building up the story of the paper. 

\begin{figure*}
	% To include a figure from a file named example.*
	% Allowable file formats are eps or ps if compiling using latex
	% or pdf, png, jpg if compiling using pdflatex
	\includegraphics[width=\textwidth]{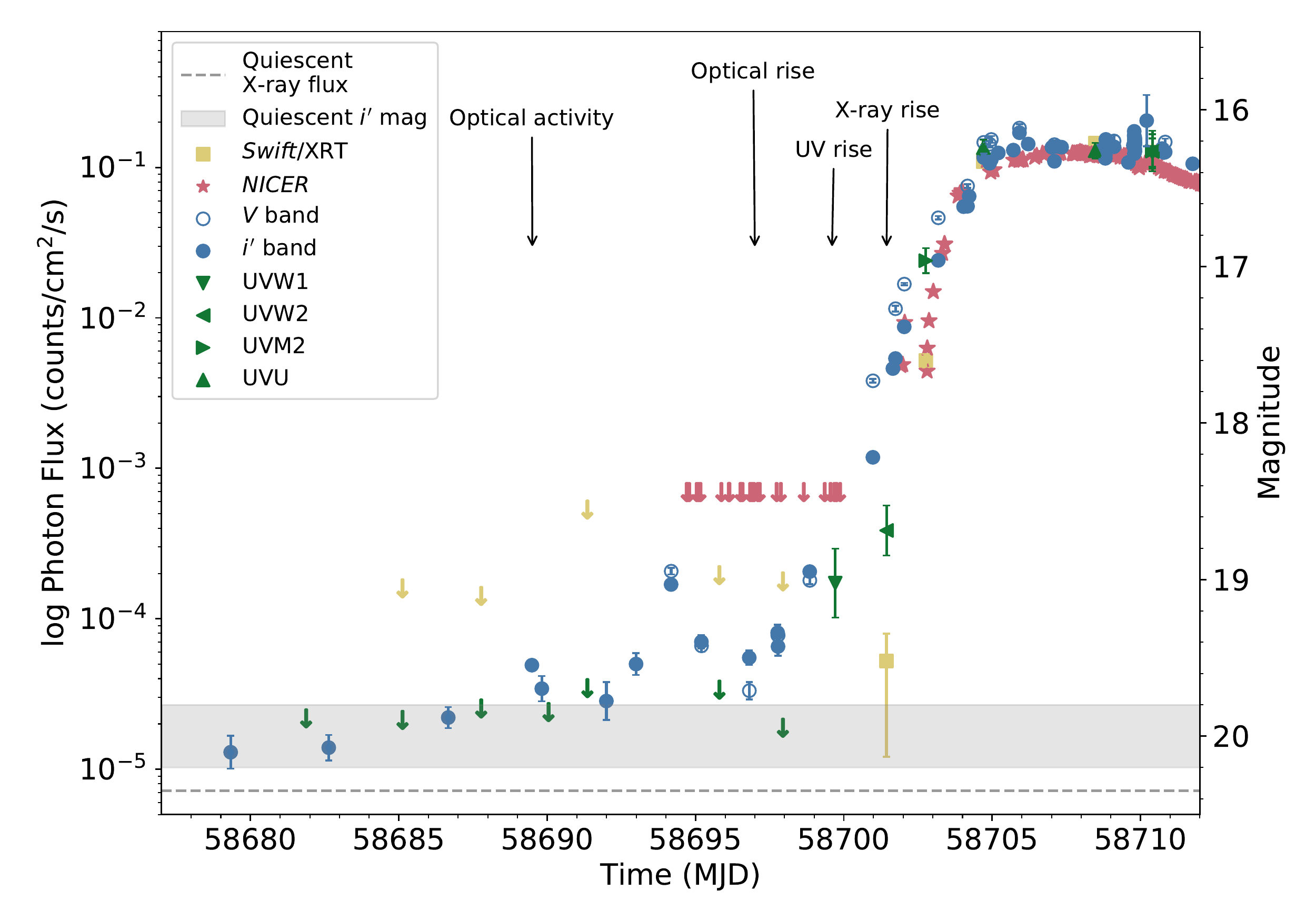}
    \caption{The multiwavelength light curve of the 2019 outburst rise in SAX J1808.4--3658. Optical $V$ and $i'$-band observations (blue circles) were obtained with the 2-m Faulkes Telescope South (at Siding Spring, Australia), and the Las Cumbres Observatory(LCO) network of 1-m robotic telescopes. \textit{Swift}/XRT 0.2--10\,keV and \textit{Swift}/UVOT UV observations are shown in yellow squares and green triangles respectively. NICER 0.2--10\,keV observations are shown in red stars. `$\downarrow$' symbols indicate UV and X-ray upper limits. Error bars and upper limits correspond to 1-$\sigma$. UV and optical observations are plotted on the RHS axis (Magnitude) and the \textit{Swift} and NICER X-ray observations are plotted on the LHS axis, in log 0.2--10\,keV counts/cm$^2$/s, where the effective area of the XRT and NICER were 
    taken as
% assumed to be 
    110\,cm$^2$ and 1900\,cm$^2$ respectively. The optical and UV magnitudes have been de-reddened.}%\A{To do: add date axis to plot, consider multiple panels as Jeroen suggests, and fix NICER upper lims}{\bf (There are a lot of overlapping data point, which makes some of the evolution hard to follow - I just suggest turning this into a two panel figure, with the optical/UV in one panel and X-ray in the other. Also, could change the arrows for the activity/rise into lines that actually cross the data? - jh).}
    \label{fig:lightcurve}
\end{figure*}

The multiwavelength light curve of the 2019 rise to outburst of SAX J1808.4--3658 is shown in Fig. \ref{fig:lightcurve}. Activity of the source was first noticed in the optical $i'$-band observations, beginning July 25 (MJD 58689), as the $i'$-band magnitude fluctuated between 0.7--1.0\,mag brighter than the quiescent level of approximately 19.8\,mag. Then, between August 3.85 and 5.98 (MJD 58698.85 and 58700.98), the $i'$-band intensity increased to approximately 2\,mag brighter than the quiescent level, and continued to increase over the coming days as the outburst progressed. The peak of the 2019 outburst was reached at around August 10 (MJD 58705) in optical, and August 14 (MJD 58709) in X-ray (Fig. \ref{fig:lightcurve}).

The first \textit{Swift}/UVOT detection occurred on August 4 (MJD 58699.71), 10 days after the first sign of increased optical activity in the source, and 2\,d before the first X-ray detection. The UV magnitude in all bands rose sharply by approximately 3\,mag over the next 4 days until the outburst peak was reached. We note that prior to August 4 there were 7 UV observations in which an upper limit only was obtained with \textit{Swift}/UVOT, which had a sensitivity of $\approx21.5$\,mag for our observations given the short exposure times of the images, and so we cannot rule out low-level UV activity prior to August 4. 

There was no detection of the source in the 8 \textit{Swift}/XRT observations between July 17--August 1 (MJD 58685--58697), with upper limits of $0.8-3\times10^{-12}$\ergpcms (0.2--10\,keV). There was a marginal detection of the source on August 2 (MJD 58697.85). However, the Aug. 2 \textit{Swift}/XRT data do not show a clear source in visual inspection, leading us to suspect a spurious detection. We note that the \textit{Swift}/XRT observations between July 17--August 4 were carried out with the XRT in WT mode, which is less sensitive than the XRT in PC mode, and thus we cannot rule out low-level X-ray activity ($<1\times10^{-12}$ \ergpcms) during this time. 

On August 6.44 (MJD 58701.44) the source was first detected with the XRT in PC mode, with a count rate of $4.55\times10^{-2}$\,count\,s$^{-1}$ in the 614\,s exposure time. 
%We used the online XRT product builder \citep{Evans2009}, which uses \textsc{xspec} to fit an absorbed power law spectrum to the lightcurve from August 6.44 (OBSID 30034129). We included a multiplicative component to model the effects of neutral absorption along the line of sight using the absorption model from \cite{Morrison1983} (model ‘tbabs’ in \textsc{xspec}). 
We estimate an unabsorbed X-ray flux (0.2--10 keV) of $2.5\times10^{-12}$\ergpcms, for a power law spectrum with photon index of 1.73 and neutral hydrogen column of $1.46\times10^{21}$\,cm$^{-2}$ \citep{DiSalvo2019}. Later that day, NICER first detected the source on Aug 6.92 (MJD 58701.92), at a 0.5--10\,keV count rate of 9.07\,count\,s$^{-1}$. For this observation and the same spectral parameters, we estimate an unabsorbed (0.5--10\,keV) flux of $(2.91\pm0.11)\times10^{-11}$\ergpcms.

This first \textit{Swift} X-ray detection on August 6 corresponds to a flux $\approx$50 times greater than the X-ray quiescent flux of SAX J1808.4--3658, of $5\times10^{-14}$\ergpcms~ \citep[0.5--10\,keV,][]{Heinke2009}. This flux is the lowest  the source has been detected in the early part of any outburst to date \citep[previously the lowest flux the source had been detected at early in outburst was in 2011 at a flux of 0.2\,mCrab, which is $\approx$100 times the quiescent X-ray flux;][]{2011ATel}. We estimate the possible observed delay between the first optical activity to the first X-ray detection above the Swift flux sensitivity limit to be 10--15\,d, based on the cadence of the optical and X-ray observations. The fact that we detected the source at such low X-ray flux likely indicates that the 12 day delay between the first optical and the first X-ray activity detected is close to the actual time it takes for the disk to completely ionise from an initial instability, and for the accretion outburst to begin.

\subsection{Optical quiescence and the first signs of activity in 2019}
\begin{figure}
	\includegraphics[width=\columnwidth]{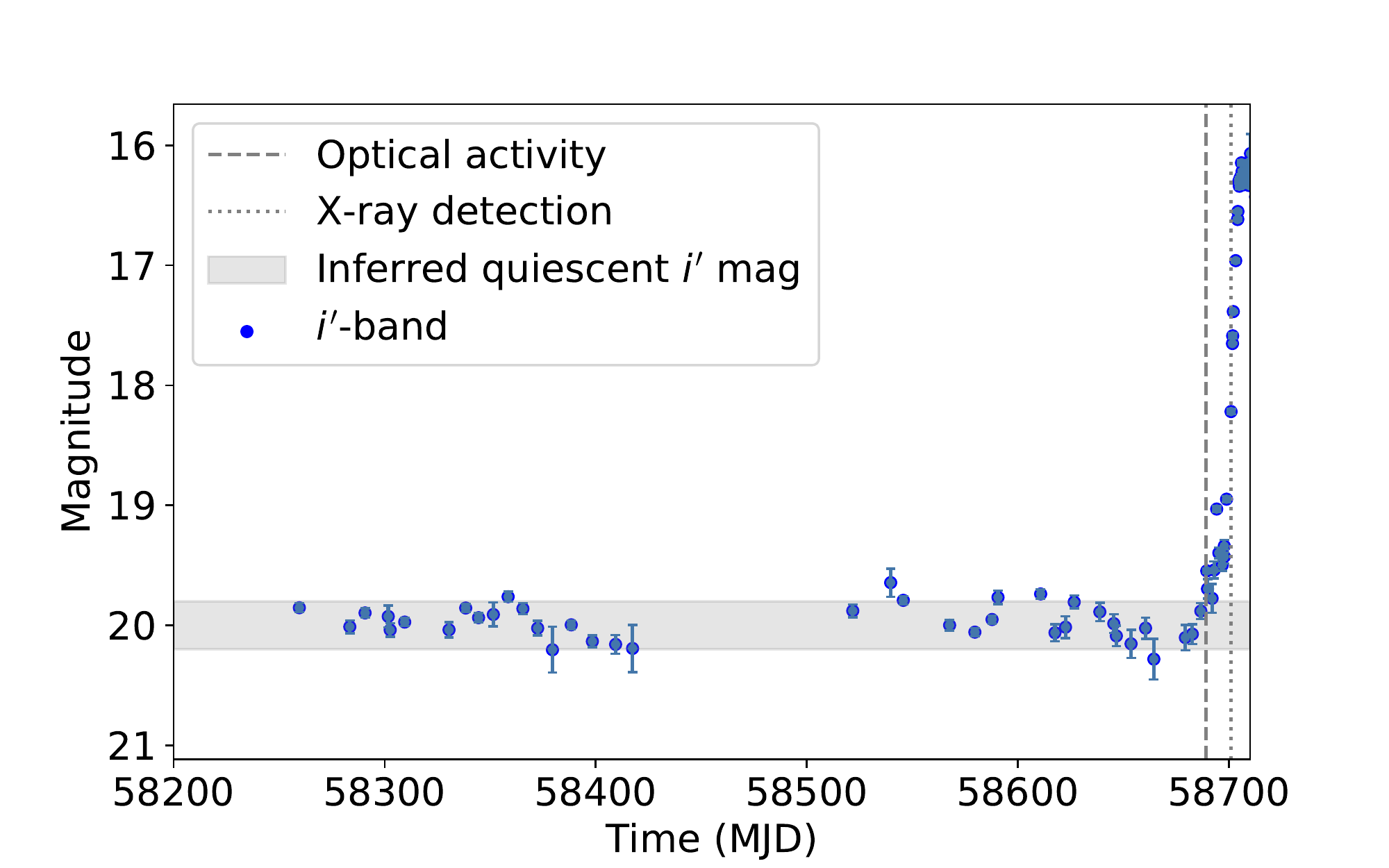}
	\caption{LCO observations of the optical $i'$-band light curve of SAX J1808.4--3658 from 500\,d before the 2019 outburst commenced. The vertical dashed line indicates the first statistically significant increase in the optical magnitude (July 25), and the vertical dotted line indicates the time of the first X-ray detection with \textit{Swift} (August 6). The horizontal grey bar indicates our inferred optical $i'$-band quiescent magnitude range, based on these observations. The magnitudes have been de-reddened.}
	\label{fig:opticalquiescence}
\end{figure}

We examined LCO $i'$-band observations of SAX J1808.4--3658 from approximately 500\,d before the 2019 outburst in order to determine the quiescent level of optical activity, and the significance of the optical activity we saw on July 25 (MJD 58689, Fig. \ref{fig:opticalquiescence}). 

The quiescent light curve shown in Fig. \ref{fig:opticalquiescence} indicates significant variations in the $i'$-band magnitude on a range of timescales. Assuming the source is in quiescence during these 500\,days, we thus measure an optical $i'$-band quiescent level of 19.8--20.2\,mag. \citet{Deloye2008} measured a de-reddened quiescent $i'$-band magnitude range of 20.05--20.55 for \sax, fainter than our inferred range. We suspect 
% this 
the discrepancy could be due to orbital variations, and confusion in the LCO observations. Alternatively, the optical quiescent level of the source could have evolved since the 2008 outburst, and the source could now appear slightly brighter in quiescence. Nevertheless, we see no evidence for optical activity above the LCO observed quiescent level prior to July 25, which is when we infer the optical activity began.

\subsection{Spectral evolution and disk temperature}\label{sec:results_spectralevolution}

\begin{figure}
	% To include a figure from a file named example.*
	% Allowable file formats are eps or ps if compiling using latex
	% or pdf, png, jpg if compiling using pdflatex
	\includegraphics[width=\columnwidth]{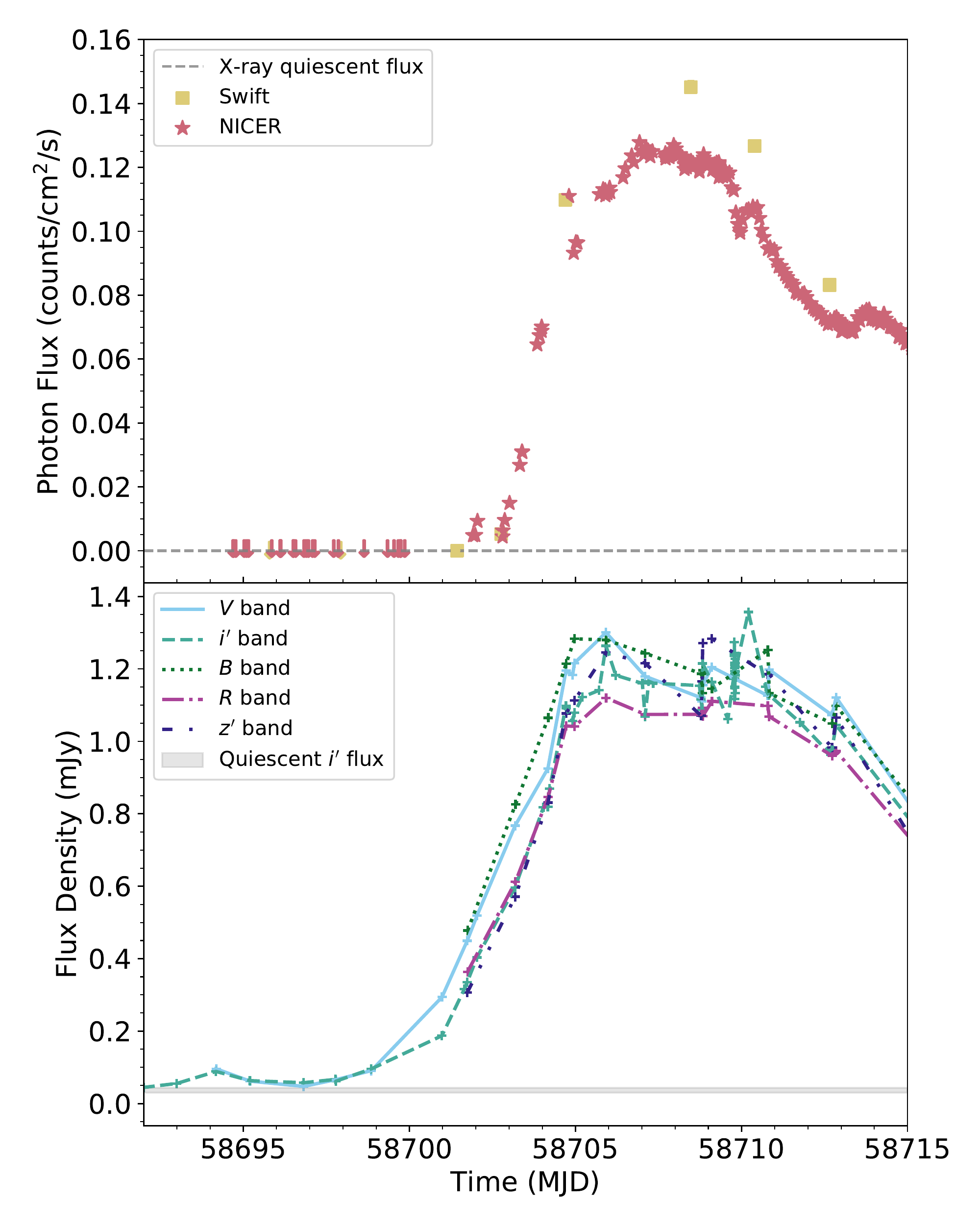}
    \caption{\textit{Bottom:} The linear X-ray and optical light curves of the 2019 outburst rise of SAX J1808.4--3658. Optical $V$, $i'$, $B$, $R$, and $z$-band observations were obtained with the 2-m Faulkes Telescope South (at Siding Spring, Australia), and the Las Cumbres Observatory (LCO) network of 1-m robotic telescopes. `$+$' symbols indicate time of observations. The optical magnitudes have been de-reddened.
    \textit{Top:} \textit{Swift}/XRT 0.2--10\,keV and NICER 0.2--10\,keV observations are shown in yellow squares and red stars. '$\downarrow$' symbols indicate X-ray upper limits. Error bars and upper limits correspond to 1-$\sigma$. The \textit{Swift} and NICER X-ray observations are plotted in 0.2--10\,keV counts/cm$^2$/s, where the effective areas of the XRT and of  NICER were 
    taken as 110\,cm$^2$ and 1900\,cm$^2$ respectively. The optical observations indicate the $V$-band rose before the $i'$-band, and the fastest increase was seen in the $B$-band.}%\A{To do: add date axis to plot, consider multiple panels as Jeroen suggests, and fix NICER upper lims}{\bf (There are a lot of overlapping data point, which makes some of the evolution hard to follow - I just suggest turning this into a two panel figure, with the optical/UV in one panel and X-ray in the other. Also, could change the arrows for the activity/rise into lines that actually cross the data? - jh).}
    \label{fig:linearlightcurve}
\end{figure}

The linearised optical and X-ray light curves are shown in Fig. \ref{fig:linearlightcurve}. It is clear the $V$-band rose more quickly than and before the $i'$-band, with limited early observations in the $B$, $R$, and $z$ bands. The $B$-band appears to rise the quickest, and reach the peak earlier than the other bands. Interestingly, the $V$-band was also observed to rise first and the $B$-band was observed to rise the fastest in observations of the dwarf Nova WX Hydri \citep{Kuulkers1991}. In the 1996 outburst of GRO J1655--40, the $B$-band was observed to rise after the $V$, $R$ and $i$ bands \citep{Hameury1997}. %B-V-i'-z-R

\begin{figure}
	% To include a figure from a file named example.*
	% Allowable file formats are eps or ps if compiling using latex
	% or pdf, png, jpg if compiling using pdflatex
	\includegraphics[width=\columnwidth]{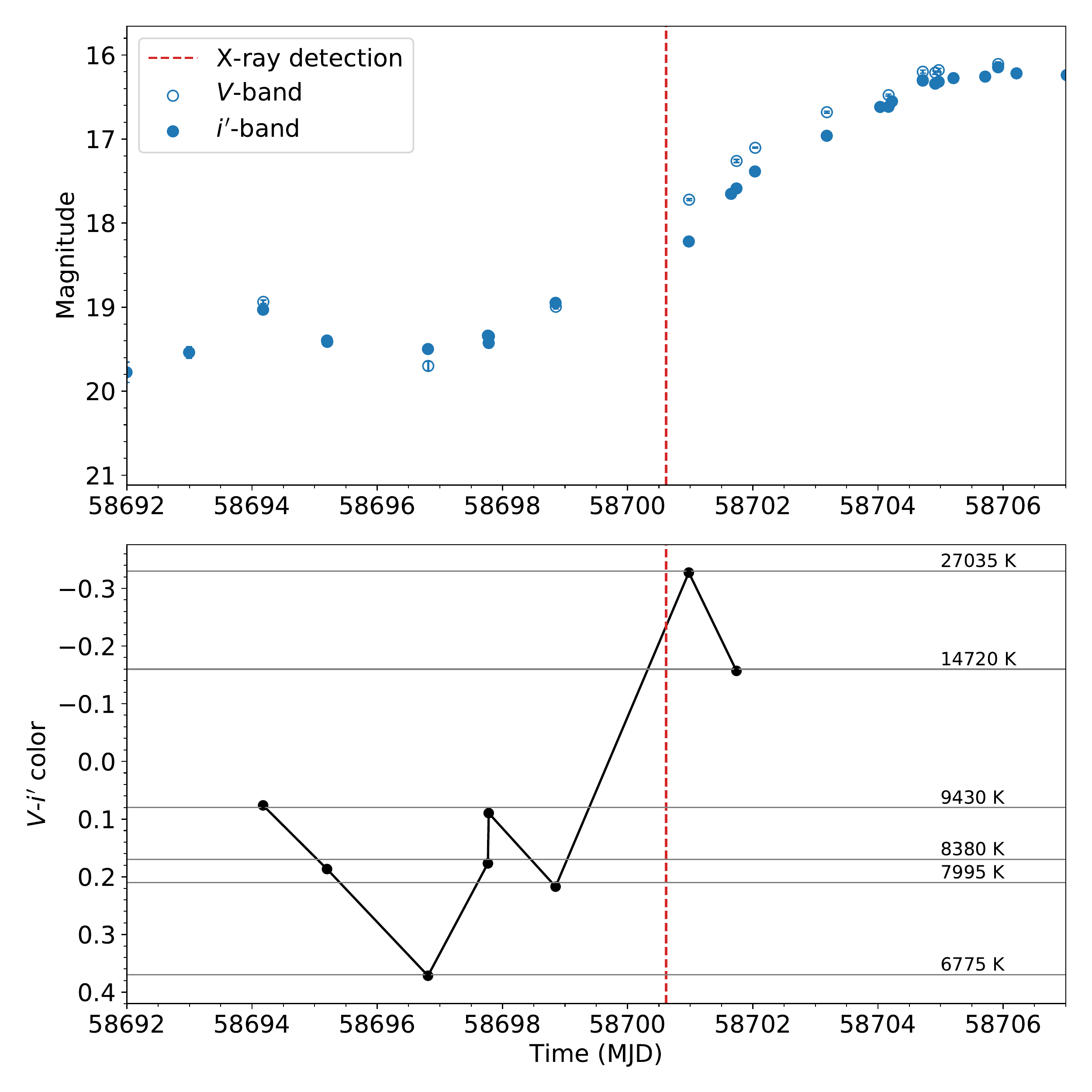}
    \caption{\textit{Top:} Optical $V$- and $i'$-band observations of the 2019 outburst of SAX J1808.4--3658 obtained with the LCO.
    \textit{Bottom:} The evolution of the $V$-$i'$ colour during the rise to the 2019 outburst of SAX J1808.4--3658. The grey horizontal lines correspond to the temperature (K) inferred from our simple blackbody disk model.
    The vertical red line indicates the time of the first \textit{Swift} X-ray detection. Statistical error bars are plotted for the $V$-$i'$ colour, but are too small to see, and do not take into account systematic uncertainties such as orbital modulations. }
    \label{fig:opticalcolor}
\end{figure}

We determined an optical 2-band $V$-$i'$ temperature using the $V$ and $i'$-band LCO observations, shown in Figure \ref{fig:opticalcolor}. To obtain a temperature from the $V$-$i'$ colour, we approximate the optically emitting part of the disk by a blackbody with time-variable temperature to model the relationship between colour and magnitude. We adopted the model of \citet{Maitra2008}, described in detail in \citet{Russell2011}. The blackbody model depends on the intrinsic colour, and thus the assumed extinction.

%as well as the apparent size of the blackbody, or the accretion disk radius. We estimate the accretion disk radius to be equal to the Roche Lobe radius of $3.87\times10^{10}$\,cm \A{Dave- is this right?},
%\A{Dave R to fill in}\,cm,
%  Roche radius = 1.29 light seconds
% # The relation between observed colour and
% # intrinsic spectral slope (alpha) in this
% # case is:
% #
% #   alpha(color) = -2.828 * color + (0.767)
%based on the known orbital period of 2.01\,hr, the NS mass of 1.4\,\msun, the companion star mass of 0.05\,\msun \citep{Chakrabarty1998}, the distance of 3.5\,kpc \citep{Galloway2006}, and inclination of 68\,degrees \citep{GoodwinMCMC}. 
How well the blackbody model fits the data depends on the uncertainties in the system parameters, as well as the disk geometry, so we do not fit the model to the data but use a normalisation parameter to approximately overlap the model with the data. This approach
% enables an 
allows us to estimate the approximate disk temperature,
% to be obtained, 
and to infer the general trend in the 
% change in disk 
temperature during the onset of outburst. % to be inferred. 

We observed the disk transitioning through the temperature at which hydrogen ionises in the 12\,d of optical activity, before X-ray emission was observed. In Figure \ref{fig:opticalcolor} the disk begins at a temperature of $\approx$10,000\,K, 5 days after the first sign of increased optical activity on July 25, and decreases in temperature to $\approx$7,000\,K on Aug 2, then spends approximately 2\,d at a temperature of $\approx8,000-9,500$\,K, and finally increases to a temperature of $>15,000$\,K, coincident with the first X-ray detection of the source on Aug 6. After Aug 6, the $V$-$i'$ band colour begins to increase again, and our simple blackbody disk model breaks down as the optical spectrum now likely has contributions from the outer irradiated disk as well as a jet component (Baglio et al., submitted). Thus, after Aug 6 we cannot infer the disk temperature using our model, so it is unlikely that the accretion disk temperature began decreasing again, and more likely that the jet component of the optical emission is contributing to the $V$-$i'$ colour, which our disk model does not include. 

We note that we do not have exactly simultaneous $V$ and $i'$ magnitude measurements to infer the temperatures in Fig. \ref{fig:opticalcolor} from the $V$-$i'$ colour, which would cause larger uncertainties in temperature at the brighter magnitudes, or hotter temperatures. The fainter magnitude points earlier in the rise are less affected by this uncertainty, as 0.1\,mag corresponds to $<1000$\,K at these lower fluxes. The error bars we plot are statistical only, and do not take into account systematic uncertainties in the $V$-$i'$ colour. The colour-magnitude diagram of the whole outburst, and the model fit of the data, are presented in (Baglio et al., submitted).

The  range of temperatures we inferred from the beginning of optical activity to the first X-ray detection corresponds to the range of ionisation temperatures of hydrogen, with hydrogen likely being completely neutral below 5,000\,K, and completely ionised at 10,000\,K \citep*{Lasota2008}. This observation is crucial in confirming the role of hydrogen ionisation in the DIM, and triggering outbursts in LMXBs.

\subsection{Optical spectrum}
\begin{figure*}
	% To include a figure from a file named example.*
	% Allowable file formats are eps or ps if compiling using latex
	% or pdf, png, jpg if compiling using pdflatex
	\includegraphics[width=\textwidth]{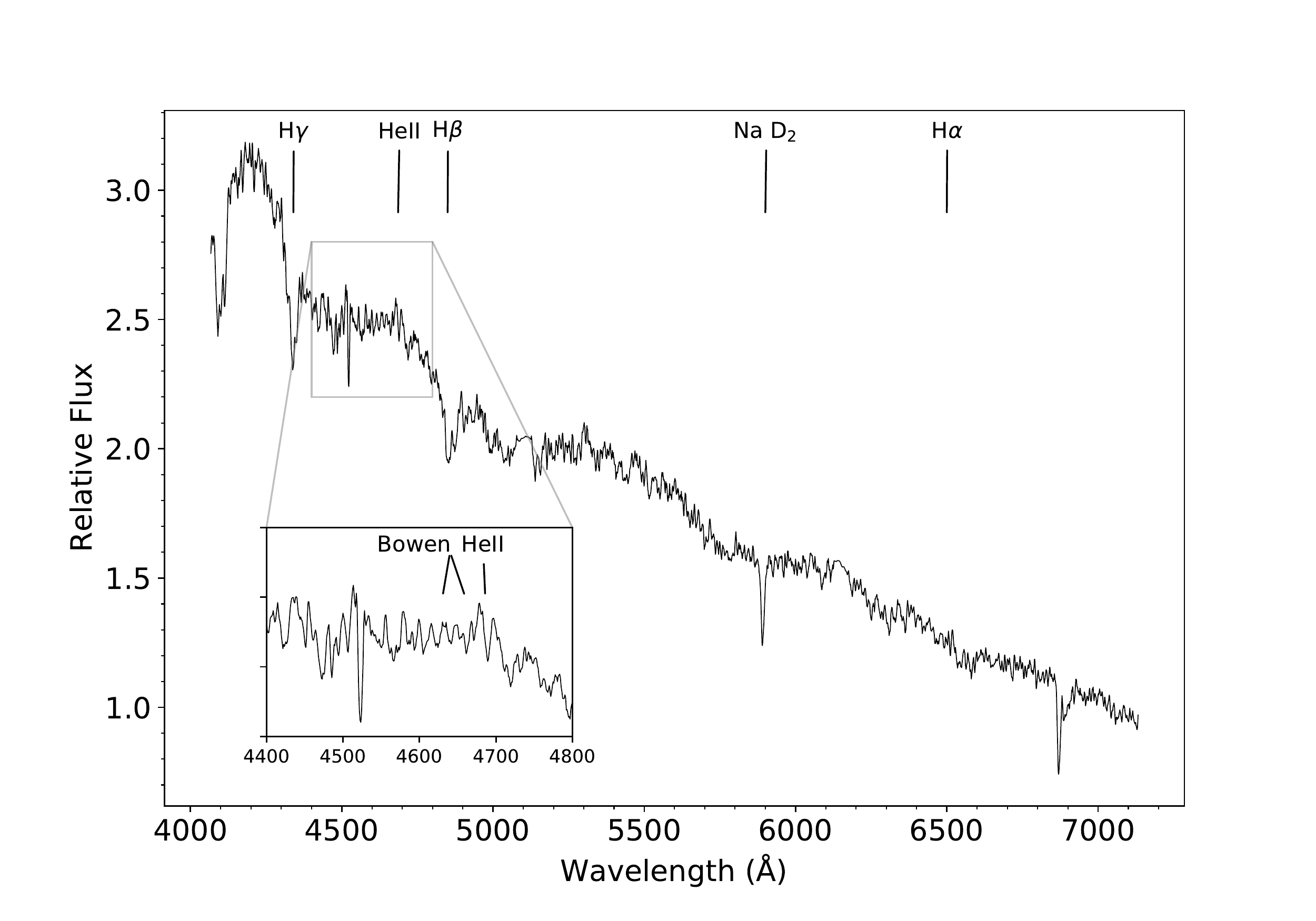}
    \caption{SALT summed (2400\,s) optical spectrum of SAX J1808.4--3658 taken on 2019 August 6 (MJD 58701). There are clear Balmer absorption features and weak HeII emission present.}
    \label{fig:saltspectra}
\end{figure*}

The summed optical spectrum from 2019 August 6 (MJD 58701), shown in Figure \ref{fig:saltspectra}, is consistent with an optically thick accretion disk viewed at low-intermediate inclination. The spectrum is similar to that taken by \citet{Cornelisse2009} during the 2008 outburst, with strong H$\alpha$, H$\beta$, and H$\gamma$ absorption lines, but with weak HeII emission and no Bowen complex emission visible. Central emission reversals are evident in the H$\alpha$ and H$\beta$ absorption lines, almost filling in the former.  A strong HeII emission line was observed later in the outburst in additional spectra obtained with SALT (Russell et al., in preparation), and is present in the spectrum obtained by \citet{Cornelisse2009} during the 2008 outburst, but no Bowen complex emission was detected in any of the spectra obtained with SALT during this outburst. 

The accretion disk in SAX J1808.4--3658 is likely composed of $\approx$50\,$\%$ helium, inferred from properties of the type I X-ray bursts observed from the source \citep{GoodwinMCMC,Galloway2006,Johnston2018}. The spectrum observed on Aug 6 coincides with an inferred accretion disk temperature of $\approx$10,000\,K (Section \ref{sec:results_spectralevolution}), right at the temperature that helium in the disk would begin to be ionised. The fact that only weak HeII emission was observed in the spectrum, and much stronger HeII emission was observed later in the outburst implies that the accretion disk was not yet hot enough at this time to have ionised much of the helium in the disk, and later in the outburst more helium in the disk was ionised.

\section{Discussion}\label{sec:discussion}
% This is where you evaluate your data and discuss the implications
% Answer: Why should I trust your data? Why should I care?

% start by helping readers regain focus, with a brief statement of the culminating feature of the results ("Our findings reveal..")

% At some point after this, examine the findings from the viewpoint of a reviewer, and anticipate their most important comments. 

% Set the most important findings into context with the literature. Usually picking up the narrative in your introduction again. What degree do your findings move the field forward? 
% Does it help resolve a controversy? Are there any aspects still unresolved?
% Are there any features that couldn't be explained? If so, what might that be telling us?

Our findings reveal a 10--15\,d delay between the optical and X-ray activity at the onset of outburst in the accretion-powered millisecond pulsar SAX J1808.4--3658, as well as a 4\,d optical to X-ray rise delay. This work presents
%delay is the longest 
% delay between optical and X-ray brightening 
%ever observed in an LMXB, and 
the first observational measurement of the X-ray delay time in an LMXB using an X-ray instrument more sensitive than an all sky monitor, and provides important observational constraints for the disk instability model. Our first X-ray detection at $\approx$50 times the quiescent X-ray level confirms that X-ray activity before this detection was either non-existent or very low level. The 12\,d delay (possible range of 10--15\,d) between the X-ray and optical activity we observed is indicative of the time between the initial beginnings of activity in the system, to the triggering of the outburst through a disk instability, to the onset of accretion onto the neutron star.

SAX J1808.4--3658 is a very well-studied source, and has precisely measured orbital constraints, listed in Section \ref{sec:introduction}. These observational constraints imply the accretion disk in SAX J1808.4--3658 is small ($R<3\times10^{10}$\,cm), and the inner disk is likely truncated by the neutron star magnetic field \citep{Papitto2009} at the neutron star magnetospheric radius \citep[see e.g.][]{Frank2002}. These are important considerations to take into account when assessing the possibility of the outburst being explained by the DIM.

There was no radio detection coincident with the position of \sax\ before the first X-ray detection on August 6, in an observation by MeerKAT on July 31 \citep{MeerkatAtel}, but there was a radio detection by MeerKAT at approximately the peak of the outburst, on August 10 \citep{MeerkatAtel2}. Since radio emission is generally thought to come from a jet component of the system, 
% this would indicate that 
the nondetection on July 31 
% there 
indicates that there was not yet a jet present, which would be consistent with the disk not yet filling in the truncated inner edge to reach the neutron star. In addition, the first indication of a jet at optical wavelengths (i.e. when the jet power was relatively high) occurred after August 8 (Baglio et al., submitted).

The temperature of the accretion disk we inferred during the optical rise, of 7000\,K increasing to $\gtrsim20000$\,K at the time of the first X-ray detection, corresponds exactly to the ionisation temperature range of hydrogen. Furthermore, in the spectrum obtained on August 6 (Figure \ref{fig:saltspectra}), the strong Balmer absorption lines present are a clear indication that hydrogen was ionised in the disk. These inferred temperatures and level of hydrogen ionisation thus support the prevalent theory of the disk instability model: that the initial instability that triggers an outburst is caused by ionisation of hydrogen in the accretion disk \citep[e.g.][]{Lasota2001,Dubus2001}. 
\subsection{The disk instability model}
We now discuss whether the observations we present are consistent with the DIM theory, and if there is evidence for an outside-in or inside-out type outburst. Classically, there is a general understanding that there would be no X-ray delay observed for an inside-out type outburst, and there would be an X-ray delay for an outside-in type outburst. However, \citet{Menou1999} and \citet{Lasota2001} are careful in asserting that an X-ray delay observed during an LMXB rise to outburst is not evidence of an outside-in outburst, as this delay is theoretically possible for an inside-out type outburst if the disk is truncated. In this scenario, ignition would occur at the truncated inner edge of the accretion disk, which is still far from the compact object, but technically is classified as an inside-out type outburst. Theoretically, inside-out outbursts are expected for LMXBs, primarily due to the slower rise times, larger disks, and the observed shape of the outburst light curves in these systems \citep{Lasota2001}. However, due to a lack of comprehensive multiwavelength observations of the rise to outburst of LMXBs, there is no compelling evidence that they exhibit either outside-in or inside-out outbursts.

The most important determining parameters of the location of the maximum surface density in the accretion disk, and thus the type of outburst that will occur in an LMXB are the mass transfer rate to the disk, the size of the accretion disk, and if the disk is truncated. According to \citet{Lasota2001}, the mass transfer rate criterion for outside-in outbursts to occur is given by,

\begin{equation}\label{eq:criticalmdot}
    \dot{M}_{\mathrm{tr}} \gtrsim \dot{M}_\mathrm{A}
    = 
    2\times10^{15}\delta^{-1/2}
    \left(\frac{M_{\mathrm{NS}}}{M_{\mathrm{\odot}}}\right)^{-0.85} 
    \left(\frac{R}{10^{10} \mathrm{cm}}\right)^{2.65} 
    \quad \mathrm{g\,s^{-1}}
\end{equation}

where typically $\delta\leq2$. 

Thus, for SAX J1808.4--3658, where the quiescent accretion rate has been inferred to be $5.7\times10^{14}$\,g\,s$^{-1}$ \citep{Heinke2009}, the critical mass transfer rate for an outside-in outburst to occur is reached if the disk is smaller than $\approx7\times10^9$\,cm. However, this criterion was developed for a CV system, and while it is likely that this criterion applies equally well to a neutron star LMXB such as SAX J1808.4--3658, there may be some important differences. Including irradiation (or heating of the outer disk) lowers the critical mass transfer rate due to additional heating of the outer disk region, lowering the maximum surface density, shortening the distance between the maximum and minimum surface densities, and allowing an outside-in outburst to occur for a slightly larger disk for SAX J1808.4--3658 \citep{Lasota2001}. A thermal instability is easier to be triggered close to the outer disk radius when irradiation is accounted for. 
%The Roche Lobe in SAX J1808.4--3658 has a radius of approximately $3.87\times10^{10}$cm, which places a maximum bound on the radius of the disk. 
Thus, the system parameters in SAX J1808.4--3658 appear to be quite similar to the critical mass transfer rate for an outside-in outburst to occur, and since it is not clear how much irradiation might affect these calculations, we cannot deduce if the system theoretically should exhibit inside-out or outside-in type outbursts from Equation \ref{eq:criticalmdot} alone.

There are a few different observationally constrained rise times important to consider when understanding the occurrences in the disk that lead to this outburst. First, the optical timescale, from the initial optical brightening on July 25 (MJD 58689), to the clear beginning of the optical rise on August 1 (MJD 58696), to the peak of the optical light curve on August 10 (MJD 58705), indicates a maximum optical rise time of 16 days for the disk in SAX J1808.4--3658, but it is more likely the true optical rise time is shorter than this (closer to 10\,d) and the low-level optical activity between July 25 and August 1 is due to precursor activity before the onset of the outburst. Next, the UV delay time, between the initial optical activity on July 25 to the first UV detection above the sensitivity of \textit{Swift}/UVOT on August 4 (MJD 58699), of 10\,d, as well as the optical to UV rise delay time of 2\,d. Next, the X-ray rise time from the first X-ray detection on August 6 (MJD 58701) to the peak of the X-ray light curve on August 14 (MJD 58709), of 8 days. Finally, the X-ray delay, between the commencement of the optical rise and the X-ray rise of 4\,d, and between the optical brightening and the X-ray brightening, of 12 days. We note that these inferred timescales are heavily dependent on the sensitivity of the instruments used, and there is the possibility that there was low level activity in UV and X-ray before statistically significant detections were obtained.

\subsection{The viscous timescale}
According to \citet{Lasota2001}, the UV and X-ray flux will rise when the temperature has grown sufficiently high, when the heating front arrives at the inner disk. One may expect the inner disk to arrive close to the neutron star's surface in the viscous time, which is given by,

\begin{equation}
    t_{\mathrm{visc}} = \frac{R\Delta R}{\nu}
\end{equation}
where $R$ is the inner disk radius (i.e. the truncation radius), $\Delta R$ is the typical scale of the density gradient, $\Delta R\approx\sqrt{HR}$, and $\nu$ is the viscosity, which is described by the $\alpha$-prescription \citep{Shakura1973},

\begin{equation}\label{eq:tvisc}
    \nu = \alpha \mathrm{c_s} \mathrm{H}
\end{equation}
where $\alpha$ is the artificial viscosity coefficient, $c_s$ is the sound speed, and $H$ is the scale height. 

For SAX J1808.4--3658, the inner disk radius is less than the Roche lobe radius of $3.87\times10^{10}$\,cm, the neutron star mass is $\approx1.4$\msun, the neutron star radius is $\approx11.2$\,km, and the accretion rate into the disk in quiescence is $\dot{M}_{tr}=5.7\times10^{14}$\,g\,s$^{-1}$ \citep{Heinke2009}. Using the Shakura-Sunyaev disk solution outlined in \citet{Frank2002} and assuming an inner disk radius of $5\times10^8 < R < 2\times10^{10}$\,cm, in order for the viscous time of the disk to be $\approx12$\,d, $0.0009 < \alpha<0.1$. For this disk solution we find $H/R\approx0.01$ and $c_{\rm{s}}\approx20$\,km/s. Alternatively, for a viscous time scale of $\approx4$\,d, which is coincident with the delay between the first X-ray detection and the beginning of the optical rise we observed, we infer $0.0035 < \alpha < 0.9$ for the same disk size. \citet{Shakura1973} assert that $\alpha\leq1$, \citet{Tetarenko2018} inferred that $\alpha\sim 0.2-1$ in models of LMXBs, and that the typical range for a hot disk is $\alpha \approx 0.1-1$. 
%We find the 4\,d delay is compatible with the inner disk radius starting from $2\times10^9$\,cm (if $\alpha=0.1$), up to the disk size of $2\times10^{10}$\,cm (if $\alpha=0.5$). 
We find a reasonable range for $\alpha$ and the truncation radius of the disk if the viscous timescale is coincident with the 4\,d day delay. If a 12\,d day delay is assumed, we find a very limited range of alpha and disk truncation radii ($R\approx1\times10^{10}$ and $\alpha \approx 0.1$). Thus this inferred artificial viscosity coefficient suggests alternative activity in the system during the 8\,d of extended optical activity in the lead up to the outburst. 

%The range here for $t_{\mathrm{visc}}=12$\,d is not consistent with the \textbf{estimates} of \citet{Tetarenko2018}, but is consistent with other studies, including those that assume a 2-$\alpha$ model \citep[e.g.][]{Menou2000}.

% There are two unknowns here:  the truncation radius and alpha. Hameury et al. assumed a typical alpha and derived R. The authors chose to explore a range of truncation radius from 5e8 cm to 1e10 cm and derive alpha. I calculate this leads to 0.03<alpha<0.3 with a 4 day delay. For comparison, the typical value of alpha for a hot disc is 0.1 to 1 (e.g. Tetarenko et al. etc). I find it clearer to say that the delay is compatible with the inner disc starting somewhere from 2e9 cm (if alpha =0.1) up to the disc size of 2e10 cm  (if alpha=0.5, in this case the disc is basically just a narrow ring). My conclusion would be that there is a reasonable range in alpha and truncation radius that can account for the 4 day delay. This range is very limited if a 12 day delay is assumed  (around 1e10cm for alpha=0.1) but, as I said in my previous report, I do not think this is the relevant timescale to use here.

The viscous timescale has been inferred in a few black hole LMXBs, all with much longer orbital periods than SAX J1808.4--3658, from measurements of X-ray delays later in the outbursts. These include: LMC X--3, of 5--14\,d \citep{Torpin2017,Steiner2014,Brocksopp2001}; GX 339--4, of 15--20\,d 

\noindent\citep{Homan2005}; 4U 1957+11, of 2--14\,d \citep{Russell2010}; and Swift 1910.2--0546, of $\approx6$\,d \citep{Degenaar2014}. Since these systems likely have much larger disks than in SAX J1808.4--3658, and the viscous time depends so heavily on disk radius (Eq. \ref{eq:tvisc}), it is unlikely that the viscous timescale for SAX J1808.4--3658 is as long as 12\,d.

%We infer from these different time scales that it takes $\approx$12\,d for the inner disk edge in SAX J1808.4--3658 to reach the neutron star, and for high energy X-ray emission to begin.  
\subsection{The optical precursor}
We observed the optical magnitude fluctuating by $\sim$1\,mag around July 29 (MJD 58693), before the onset of the clear optical rise of the outburst on August 1. Since it is unlikely that the viscous timescale in SAX J1808.4--3658 is as long as 12\,d, the 8\,d of increased optical activity we observed between July 25--August 1 is unlikely to be due to heating fronts propagating in the disk, and more likely can be attributed to other activity in the system in the leadup to outburst. This activity could include a number of different phenomena.

The first possibility is that there was enhanced mass transfer from the companion, which would help trigger the outburst. With no (or low-level) X-ray activity during this period, there is no evidence of increased mass transfer onto the neutron star, but this does not rule out increased mass transfer to the disk. 

Another explanation is that the emission could have arose from instabilities due to geometric effects in the outer disc, involving spiral waves and/or the accretion stream impact point, interacting with the increasing density of the outer disk. This could lead to enhanced optical activity from these instabilities in the outer disk, and eventually leading to heating fronts propagating through the entire disk, and commencing the outburst. 

Alternatively, it could have been due to a small unstable branch in the thermal stability curve as the density was close to, but not quite at, the critical density at which the outburst could begin in full \citep[e.g.][Fig. 7]{Menou2000}. However, this kind of unstable activity would occur at lower temperatures than the ionisation temperature of hydrogen, and we observed the disk to be above 6,000\,K during this period (Fig. \ref{fig:opticalcolor}). 

An additional possible scenario is that the level of irradiation of the companion was fluctuating caused by changes in the pulsar radiation pressure. It has been proposed that during quiescence SAX J1808.4--3658 switches on as a radio pulsar \citep[e.g.][]{Burderi2003}, and the optical emission during quiescence is due to reprocessed emission from the companion being heated by the pulsar wind. However, there is no physical motivation for the pulsar wind to fluctuate on this kind of timescale.
Nevertheless, SAX J1808.4--3658 has been proposed to be a "hidden" black widow system \citep{diSalvo2008}, in which it is possible that the accretion flow from the companion is swept away by the pulsar radiation pressure,
%shocks off the transferred material\textbf{,
ejecting a significant amount of mass at the inner Lagrangian point during quiescence. This shock emission would then irradiate the companion star, which would reprocess the shock emission photons and emit them in the optical band. Since the shock properties, and thus level of irradiation of the companion star, depend on the pulsar wind and the mass flow, fluctuations could be observed. \citet{Burderi2003} inferred the temperature of the companion star to be approximately 5500\,K, which is similar to the temperature of the optical emission we measured of 6,000--7,000\,K, so it is not impossible that the optical fluctuations were caused by fluctuations in the level of irradiation of the companion star. This scenario could lead to fluctuations in the inner disk radius, which would help to trigger the outburst.
%However, it is unlikely that the optical emission observed was from the irradiated companion star due to its extremely low mass, and more likely it was from the accretion disk, with the optical $V-i'$ color indicating a temperature around 6,000--7,000\,K. 
%Thus this scenario is unlikely unless the level of irradiation of the companion also is able to cause changes in the accretion disk.

% In section 4.3 I do not think the description of the scenario by Di Salvo et al. 2008 is correct (lines 40-53), or I misunderstood what the authors meant to say. My understanding of the Di Salvo scenario is that the pulsar wind directly impinges the companion star, heating it (and triggering a wind, possibly also fed by matter blown off from the accretion disc, the latter leading to truncation). The irradiated face of the companion star is about 5500 K (Burderi et al. 2003): I don’t see this as completely off the mark from the estimated 6000 to 7000 K from the V-i’. The authors conclude « this scenario is unlikely » (line 52) but then say « likely cause [..] due to effects of the pulsar radiation pressure » (line 56) ?

We thus conclude that the most likely cause of the 8\,d optical precursor is either mass transfer fluctuations from the companion, instabilities due to geometric effects in the outer accretion disk, or due to effects of the pulsar radiation pressure.

%The prior optical activity could for instance be due to enhanced mass transfer from the companion, helping to trigger the outburst. I don’t know if this can be ruled out easily (hotspot activity would provide evidence for this). Alternatively, the activity could be due to a small unstable branch in the thermal stability curve (S-curve) that is sampled as the density increases close to the critical density at which the outburst really starts (e.g. Figure 7 in Menou et al. 2000 for an example). However, this would occur at lower temperatures than the H ionisation temperature, probably in disagreement with the observed temperature. Yet another option is that the pulsar pressure is fluctuating, such that the companion is more or less irradiated: in quiescence Burderi et al. argue that the optical is reprocessed emission with T~6e3 K due to the companion being heated by the impinging pulsar wind that is free to propagate at high latitude. This might also lead to fluctuations in the inner disc radius, helping to trigger the outburst.
\subsection{The outburst rise}

We propose that, once the thermal instabilities in the accretion disk had caused heating fronts to begin propagating, since the thin disk in this system is likely truncated during quiescence at the neutron star magnetospheric radius, the increased pressure due to the rising accretion rate in the disk pushed the magnetospheric radius inwards, and the disk extended inwards towards the neutron star. The time taken to fill in the inner depleted region of the disk is the viscous timescale, of $\approx4$\,d.
We conclude that based on the observations we obtained, it took approximately 4\,d from the initial instability in the disk for the heating fronts to propagate inwards, through the truncated disk region, and the mass accretion rate to begin rising at the inner disk. It then took a further 8\,d for the accretion rate at the inner disk to rise to its maximum rate. This scenario is similar to the models of GRO J1655--40 of \citet{Hameury1997}, in which the heating front arrives at the inner disk quickly, and propagates through the cavity of the truncated disk on the viscous time of 6\,d. 
Without detailed DIM models applicable to this system, which are outside the scope of this work, we find it difficult to conclusively deduce the type of outburst observed during the 2019 outburst. Detailed models of a system like SAX J1808.4--3658 are necessary in order to fully understand if these observations can be reconciled with the DIM, or if there is additional physics that we are missing that needs to be accounted for.

\section{Conclusions}
% Ideally, the conclusion should offer something new to readers, and a particularly effective approach can be to look beyond your research - by outlining the implications of your findings on the field and related fields, and making suggestions for future research. 
% Try to keep this short and focus on immediate implications. 

We have presented the first measurement of the X-ray delay in an LMXB using an X-ray instrument more sensitive than an all sky monitor that we are aware of, and the earliest comprehensive multiwavelength observations of the rise to outburst of an accreting neutron star system. The inferred optical disk temperature during the rise to outburst from the $V$-$i'$ color supports the DIM theory that LMXB outbursts are initiated by hydrogen ionisation in the disk. We found a delay of 12\,d between the first sign of increased optical activity in the source, and the first X-ray detection. We observed the optical rise to occur first, then the UV rise 2\,d later, then finally the X-ray rise a further 2\,d later. We interpret the 8\,d of optical activity prior to the commencement of the outburst to be due to either increased mass transfer from the companion star, geometric effects in the outer accretion disk, or fluctuations in the pulsar radiation pressure, causing changes in the irradiation of the companion star, and perhaps the inner disk radius.
%X-ray delay to be due to a hydrogen ionisation heating front propagating in the disk, which stalled due to the significant fraction of helium present, causing a cooling front to develop. We propose that another heating front developed a couple of days later, which ionised both the hydrogen and the helium in the disk, bringing the disk to the hot state and allowing 
We deduce that the viscous timescale of the disk in SAX J1808.4--3658 is approximately 4\,d, consistent with the optical to X-ray rise delay we observed. This is coincident with the time taken for the truncated disk to fill in to the surface of the neutron star, releasing UV and X-ray emission. Detailed modelling of the disk in SAX J1808.4--3658 is required in order to conclusively determine if these observations can be reconciled with the DIM, and if this source exhibits inside-out or outside-in type outbursts.  

%Further, more frequent, observations of the rise to outburst in this source are necessary to more precisely constrain the optical to X-ray delay in this system, and to obtain conclusive evidence of an ``inside-out" type outburst in SAX J1808.4--3658.  
\section*{Acknowledgements}

We thank the referee, Guillaume Dubus, for insightful and constructive comments that improved the manuscript and for contributing interesting discussion points. AJG acknowledges support by an Australian Government Research Training Program scholarship. The authors thank Daniel Price and Alexander Heger for helpful discussions, and the Swift team for approving the monitoring observations via a ToO proposal prior to the start of the outburst. 
Some of the observations reported in this paper were obtained with the Southern African Large Telescope (SALT), as part of the Large Science Programme on transients 2018-2-LSP-001 (PI: Buckley). Polish participation in SALT is funded by grant no. MNiSW DIR/WK/2016/07.
DAHB thanks the National Research Foundation (NRF) of South Africa for research funding. DMR and DMB acknowledge the support of the NYU Abu Dhabi Research Enhancement Fund under grant RE124.
DdM acknowledges funding from ASI-INAF n.2017-14-H.0 and from INAF 70/2016 and 43/2018. COH is supported by NSERC Discovery Grant RGPIN-2016-04602.
The Faulkes Telescope Project is an education partner of Las Cumbres Observatory (LCO). The Faulkes Telescopes are maintained and operated by LCO.

\section*{Data Availability}
The \textit{Swift}/XRT and UVOT observations are publicly available via the \textit{Swift} data center: \url{https://swift.gsfc.nasa.gov/sdc/}. The NICER observations are publicly available via the NICER data archive: \url{https://heasarc.gsfc.nasa.gov/docs/nicer/nicer_archive.html}. All optical data are available on request.

%%%%%%%%%%%%%%%%%%%%%%%%%%%%%%%%%%%%%%%%%%%%%%%%%%

%%%%%%%%%%%%%%%%%%%% REFERENCES %%%%%%%%%%%%%%%%%%

% The best way to enter references is to use BibTeX:

\bibliographystyle{mnras}
\bibliography{bibfile} % if your bibtex file is called example.bib

% Alternatively you could enter them by hand, like this:
% This method is tedious and prone to error if you have lots of references
%\begin{thebibliography}{99}
%\bibitem[\protect\citeauthoryear{Author}{2012}]{Author2012}
%Author A.~N., 2013, Journal of Improbable Astronomy, 1, 1
%\bibitem[\protect\citeauthoryear{Others}{2013}]{Others2013}
%Others S., 2012, Journal of Interesting Stuff, 17, 198
%\end{thebibliography}

%%%%%%%%%%%%%%%%%%%%%%%%%%%%%%%%%%%%%%%%%%%%%%%%%%

%%%%%%%%%%%%%%%%% APPENDICES %%%%%%%%%%%%%%%%%%%%%

% \appendix

% \section{Some extra material}
%%%%%%%%%%%%%%%%%%%%%%%%%%%%%%%%%%%%%%

% Don't change these lines
\bsp	% typesetting comment
\label{lastpage}
\end{document}